\documentclass[11pt]{article}

\usepackage[letterpaper,margin=1in]{geometry}
\usepackage[T1]{fontenc}
\usepackage[utf8]{inputenc}
\usepackage{lmodern}
\usepackage{graphicx}
\usepackage{multirow}
\usepackage{amsmath,amssymb,amsfonts}
\usepackage{amsthm}%
\usepackage{mathrsfs}%
\usepackage[title]{appendix}%
\usepackage{xcolor}%
\usepackage{textcomp}%
\usepackage{manyfoot}%
\usepackage{booktabs}%
\usepackage{algorithm}%
\usepackage{algorithmicx}%
\usepackage{algpseudocode}%
\usepackage{listings}%
\usepackage[numbers,sort&compress]{natbib}
\usepackage[hidelinks]{hyperref}
\usepackage{placeins}

\newcommand{\paperabstract}{}
\makeatletter
\renewcommand{\abstract}[1]{\gdef\paperabstract{#1}}
\makeatother

\theoremstyle{plain}
\theoremstyle{definition}

\theoremstyle{remark}

\begin{document}

\title{Molecular representation shapes the balance between target fidelity
and exploration in flow based polymer generation}

\author{
Tianren Zhang$^{1}$\thanks{Corresponding author: tianren@udel.edu}\\[0.5em]
\small $^{1}$Department of Materials Science and Engineering,\\
\small University of Delaware, Newark, Delaware 19716, United States
}

\date{}


\abstract{Designing polymers with targeted properties requires navigating vast chemical spaces from limited labeled data. Here we introduce PolyLatentFlow, a framework based on continuous-time flow matching in latent space for unconditional and conditional polymer generation, together with LlamaUni, a multimodal representation combining polymer sequence and 3D structural information. In unconditional generation, PolyLatentFlow with LlamaUni produced the largest yield of valid candidates novel relative to PolyInfo among the evaluated unconditional generators while maintaining high diversity. For $T_g$ conditioning, generated property distributions shifted systematically across a 200 °C target range. In multi-property tasks, molecular representations showed similar surrogate target fidelity but differed markedly in validity, training-set replay, and structural proximity to labeled polymers. PolyLatentFlow with LlamaUni consistently combined high validity with low replay and achieved the largest per-attempt yield of nonreplayed target hits for CO$_2$/N$_2$ conditioning. These results demonstrate latent space flow matching for polymer inverse design and identify molecular representation as a key determinant of target control and exploration beyond labeled chemistry.}

\maketitle

\begin{center}
\begin{minipage}{0.92\linewidth}
\small
\textbf{Abstract.} \paperabstract
\end{minipage}
\end{center}


\section{Introduction}\label{sec1}

Designing polymers with desired properties requires navigating an enormous space of possible chemistries for polymer repeat units, yet reliable experimental measurements are available for only a small fraction of this space\cite{audus2017polymer,chen2021polymer,gormley2021combinatorial,tran2024functional,mcdonald2023biomaterials}. Machine learning has substantially expanded polymer informatics, enabling rapid property prediction and screening across diverse polymer chemistries\cite{kim2018polymer,chen2020dielectric,tao2021hightemp,park2022gcn,aldeghi2022ensemble,queen2023polygnn,xu2023transpolymer,kuenneth2023polybert,qiu2024polync,zhang2025polyllmem,lin2019bigsmiles,schneider2024generativebigsmiles}. Generative models offer a route beyond screening predefined libraries and can, in principle, propose polymers not represented in existing candidate collections\cite{yue2025polybench,qiu2024polytao}. Realizing this promise for polymers, however, is constrained by the limited size of experimentally characterized datasets. Property datasets for glass transition temperature, thermal decomposition, and gas permeability typically contain only thousands of labeled polymers, with substantially fewer examples available when multiple properties are required simultaneously \cite{barnett2020gas,yang2022gas,kuenneth2021multitask}. Large virtual corpora such as PI1M provide broad structural coverage for representation learning, but consist primarily of hypothetical polymer structures rather than experimentally characterized materials \cite{ma2020pi1m,ohno2023smipoly,tiwari2024datasets}. Consequently, conditioning on target properties must still be learned from relatively small labeled datasets. A generator may therefore satisfy validity and property targets while remaining close to labeled training examples, limiting exploration beyond known polymer chemistry.

Molecular representation is a plausible determinant of the balance between control of target properties and structural exploration. A compact representation that groups polymers by features relevant to properties may facilitate conditioning from few labeled examples, but can also reduce distinctions among structurally different candidates and limit exploration beyond the labeled set. Conversely, a richer representation that preserves more structural information may support broader exploration, potentially at the expense of conditional control. Polymer informatics now employs representations ranging from conventional Morgan fingerprints \cite{rogers2010ecfp} to transformer based polymer language models \cite{kuenneth2023polybert, xu2023transpolymer}, as well as multimodal representations such as PolyLLMem \cite{zhang2025polyllmem}. However, the effect of molecular representation itself on unconditional and conditional polymer generation remains difficult to assess because existing studies typically vary representation together with model architecture, training procedure, and decoder design \cite{yue2025polybench}.

Most applications of machine learning in polymer informatics have historically focused on property prediction, while data-driven inverse design has developed through virtual screening, optimization, and increasingly direct generative approaches \cite{sattari2021inverse}. Early studies demonstrated inverse design through coarse-grained sequence optimization \cite{webb2020targeted}, experimentally evaluated design of cloud point and thermal conductivity \cite{kumar2019cloudpoint,wu2019thermal}, and reinforcement learning coupled with simulation validation \cite{ma2022reinforcement,li2026polyrl}. Generative models for polymers have since expanded from VAE and graph based approaches \cite{batra2020syntaxvae,gurnani2021polyg2g,liu2023invertible,kim2023omg,jiang2024topology,vogel2025copolymer,yue2025polybench} to transformer and conditional language models \cite{qiu2024polytao,savit2025polybart,khajeh2025conditional,sahu2026polyt5} and diffusion based frameworks \cite{yang2024electrolytes,jain2025polygen}. Continuous-time flow matching provides an alternative generative formulation that directly learns a velocity field transporting a simple prior toward the data distribution, offering flexible conditional modeling and efficient continuous-time sampling \cite{lipman2023flow}. These characteristics make latent flow matching attractive for polymer design, where broad chemical space exploration must be reconciled with property conditioning from comparatively sparse labeled data. However, its effectiveness for unconditional generation and generation conditioned on target properties, and how molecular representation influences target control, structural exploration, and reuse of the training set, remain largely unexplored.

To address these questions, we developed PolyLatentFlow, a modular framework for continuous-time flow matching in latent space for unconditional polymer generation and generation conditioned on target properties, together with LlamaUni, a multimodal representation combining Llama-3 embeddings of polymer Simplified Molecular Input Line Entry System (pSMILES) \cite{grattafiori2024llama3} with Uni-Mol embeddings of corresponding capped dimers of the repeat unit \cite{zhou2023unimol}. PolyLatentFlow maps each molecular representation into a compact latent space, learns unconditional or conditional distributions through flow matching, and reconstructs pSMILES with an autoregressive decoder. We compare LlamaUni with Morgan fingerprints \cite{rogers2010ecfp}, PolyBERT \cite{kuenneth2023polybert}, and a pSMILES-VAE using a consistent generative and evaluation framework. This design enables a controlled assessment of how molecular representation shapes polymer generation.

Across unconditional and conditional generation, PolyLatentFlow with LlamaUni combined high polymer validity, broad chemical space exploration, and low reuse of the training set. It produced the largest yield of valid and novel candidates in unconditional generation and maintained target-responsive conditional generation with competitive multi-property fidelity and substantially reduced replay. Further analyses linked these differences to behavior in latent generation and molecular decoding that depended on representation. Together, these results demonstrate the utility of latent space flow matching for polymer inverse design and identify molecular representation as a key determinant of target control and exploration beyond labeled chemistry.

\section{Results}\label{sec2}
\subsection{PolyLatentFlow pipeline}

\begin{figure}[t]
\centering
\includegraphics[width=\textwidth]{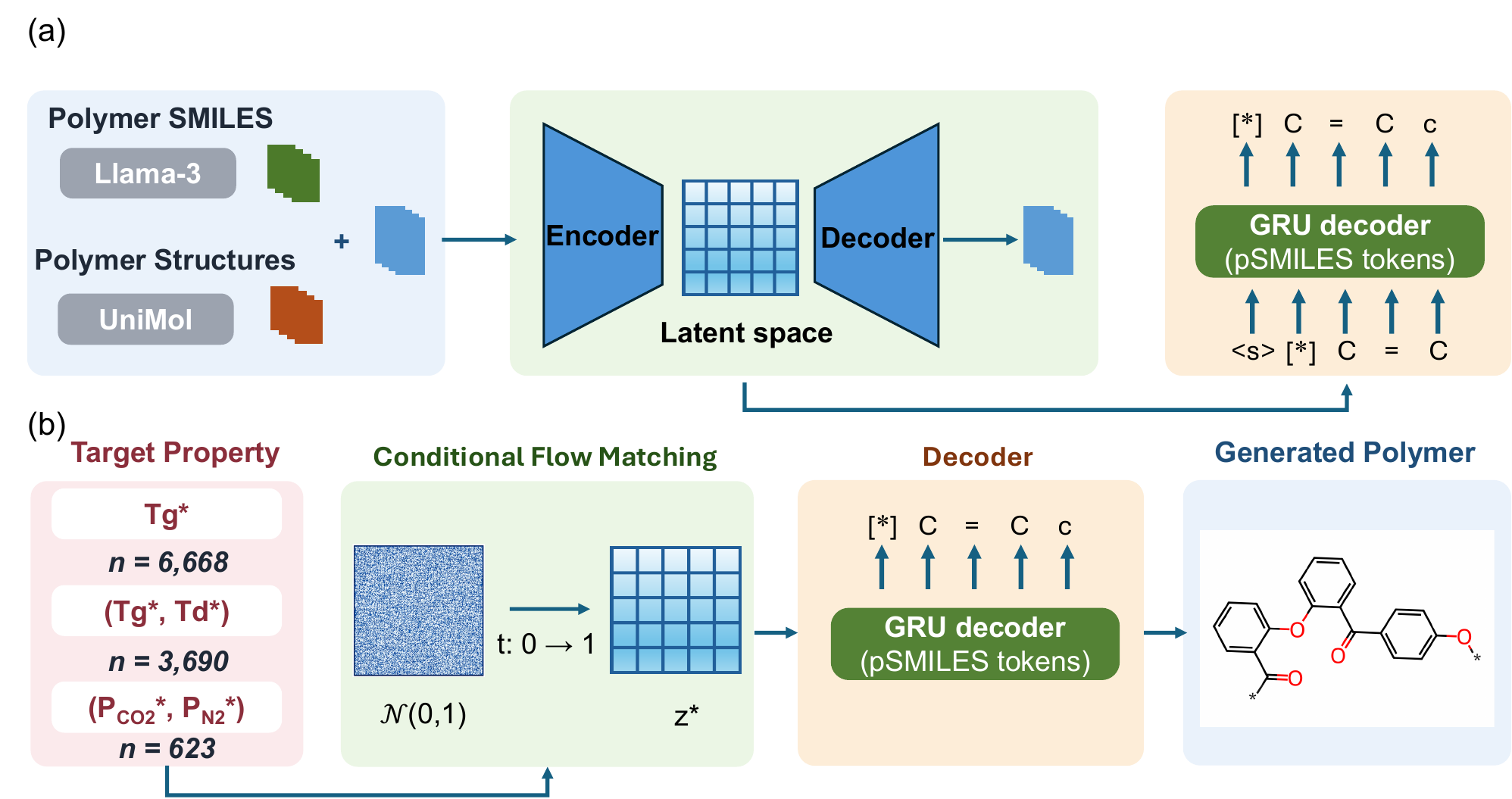}
\caption{\textbf{PolyLatentFlow workflow.} (a) LlamaUni combines a frozen Llama 3 embedding of the pSMILES with a frozen Uni-Mol embedding of the corresponding capped polymer repeat unit dimer. A VAE compresses the fused representation into a latent representation with 512 dimensions and reconstructs the fused representation. A separately trained autoregressive decoder learns to generate pSMILES token by token from the latent representation. (b) Conditional flow matching transforms Gaussian noise at \(t=0\) into a compact latent representation \(z^{*}\) at \(t=1\) conditioned on the requested property target. The frozen autoregressive decoder converts \(z^{*}\) into a pSMILES string representing the generated polymer. The datasets used for property conditioning contain 6,668 polymers with \(T_g\) measurements, 3,690 polymers with paired \(T_g\) and \(T_d\) measurements, and 623 polymers with paired CO$_2$ and N$_2$ permeability measurements.}

\label{fig:polylatentflow}
\end{figure}

We constructed PolyLatentFlow, a two stage framework that applies flow matching in latent space to unconditional and conditional polymer generation (Fig.~\ref{fig:polylatentflow}). The framework separates molecular representation, latent generation, and molecular reconstruction through a compact latent representation. Stage $1$ introduces LlamaUni, which combines complementary sequence and structural descriptions of the same repeat unit, as shown in Fig.~\ref{fig:polylatentflow}a. pSMILES are embedded using frozen Llama-3-8B, whereas capped repeat unit dimers are embedded using frozen UniMol. A variational autoencoder (VAE) is trained to compress their concatenated representation into a 512-dimensional compact latent space while reconstructing the original fused embedding. A separately trained latent-conditioned autoregressive decoder reconstructs pSMILES token by token from the compact latent. The compact VAE and decoder were trained on approximately 987,000 PI1M polymers and fixed during downstream generation. Stage 2 learns distributions over the compact latent representations using continuous-time flow matching for unconditional generation and generation conditioned on target properties (Fig.~\ref{fig:polylatentflow}b). We considered three conditioning regimes: glass transition temperature \(T_g\); joint glass transition and thermal decomposition temperatures \((T_g,T_d)\); and joint CO\(_2\)/N\(_2\) permeability \((P_{\mathrm{CO_2}},P_{\mathrm{N_2}})\). Conditional flow models were initialized from unconditional flows trained on PI1M compact latents and then fine-tuned on the corresponding labeled datasets. At inference, the requested properties guide transport from Gaussian noise to a latent representation, which the frozen decoder converts into candidate pSMILES.


To isolate the effect of molecular representation, we compared four PolyLatentFlow configurations. LlamaUni, Morgan fingerprints, and PolyBERT each used the same two stage design: their inputs were compressed into matched 512-dimensional latent spaces and decoded using separately trained Gated Recurrent Unit (GRU) decoders with identical architectures and training procedures. Latent dimensionality, flow architecture, decoder architecture, conditioning protocol, and evaluation procedure were therefore held constant across these three configurations. The fourth configuration, pSMILES-VAE, learned a direct pSMILES representation with a bidirectional GRU sequence encoder followed by a compact VAE, without a frozen upstream molecular representation. Its posterior means were supplied to flow models with the same architecture used for the other configurations and to a separately trained, representation-specific GRU decoder for generation. We therefore interpret pSMILES-VAE as an architectural comparator rather than a strictly controlled representation substitution. All architectures, training procedures, and evaluation surrogates are detailed in Methods.

\subsection{Unconditional generation}

\begin{figure}[t]
\centering
\includegraphics[width=\textwidth]{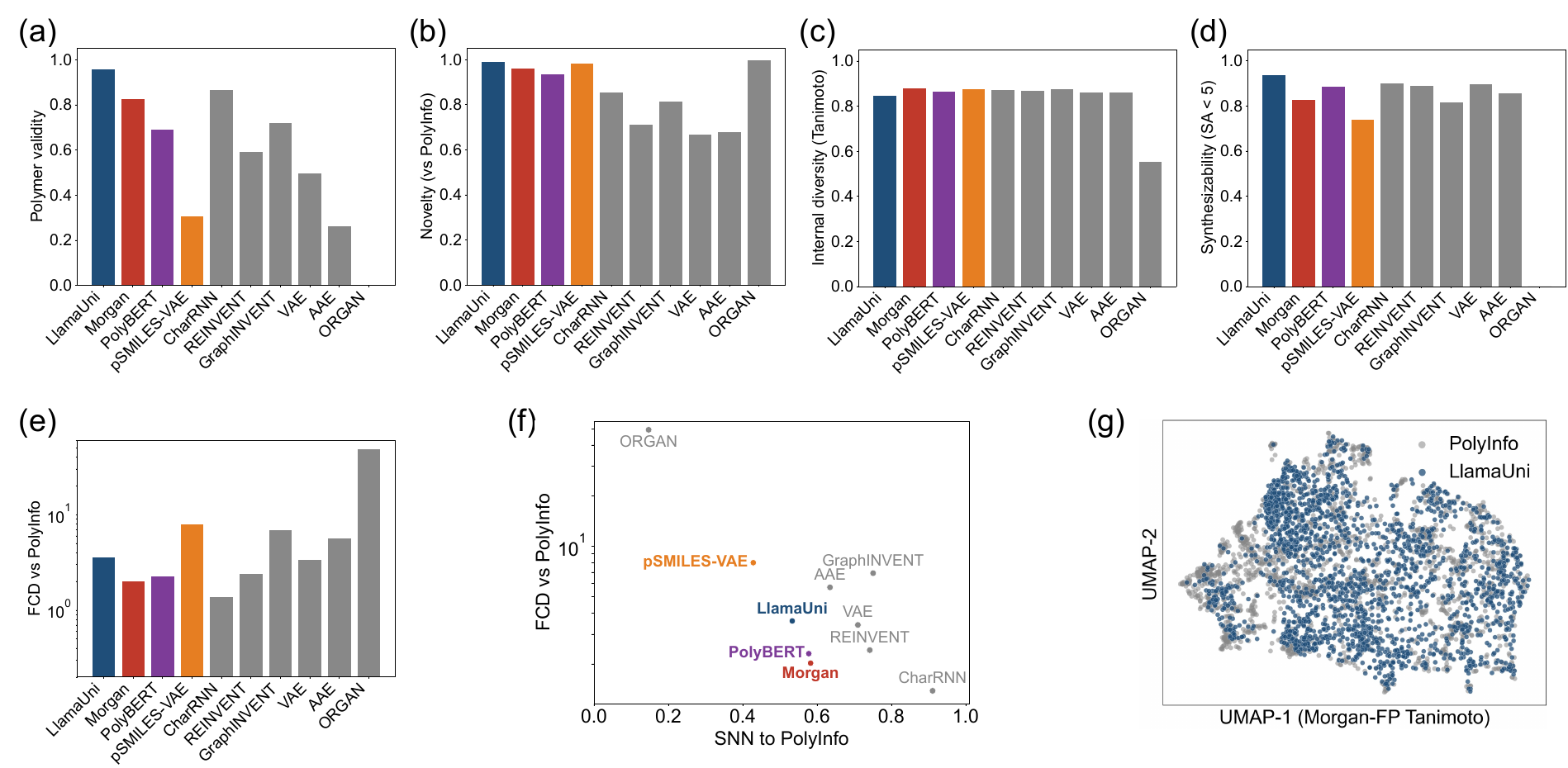}
\caption{\textbf{LlamaUni produces the largest yield of valid polymers novel relative to PolyInfo while maintaining broad chemical space coverage.} (a) Polymer validity. (b) Novelty relative to PolyInfo among unique valid polymer outputs. (c) Internal diversity calculated from pairwise Tanimoto distances using Morgan fingerprints. (d) Fraction of valid polymer outputs with an SA score below 5. (e) Fréchet ChemNet Distance (FCD) relative to PolyInfo. (f) FCD versus mean nearest neighbor similarity (SNN) to PolyInfo, comparing global distributional agreement with proximity to individual reference polymers. (g) UMAP projection in Morgan fingerprint space showing PolyInfo reference polymers and valid LlamaUni generations. Each model contributed 10,000 generation attempts.}
\label{fig:uncondgen}
\end{figure}

We generated $10{,}000$ unconditional samples from each PolyLatentFlow configuration and evaluated them alongside 10,000 samples from each of six deep generative baselines reported by Yue et al.~\cite{yue2025polybench} (CharRNN, REINVENT, GraphINVENT, VAE, AAE and ORGAN) (Fig.~\ref{fig:uncondgen}), using the same scoring pipeline. PolyLatentFlow with LlamaUni produced 9,491 valid polymer outputs absent from the PolyInfo reference set, the largest valid and novel yield among the evaluated models. This corresponded to 95.8\% polymer validity and 99.3\% novelty relative to PolyInfo among valid outputs, with validity exceeding that of CharRNN, the strongest published baseline on this metric, by 9.1\%. The other PolyLatentFlow configurations also maintained high novelty relative to PolyInfo, whereas ORGAN's apparent 100\% novelty was not considered competitive because only 0.4\% of its outputs were valid polymers (Fig.~\ref{fig:uncondgen}a,b).

LlamaUni achieved a high yield of valid and PolyInfo novel polymers without an evident loss of output diversity. Internal diversity remained between 0.83 and 0.88 across the PolyLatentFlow configurations and was comparable to that of the published baselines (Fig.~\ref{fig:uncondgen}c and Table S1). LlamaUni also produced the largest fraction of valid outputs satisfying the SA score threshold of 5 (Fig.~\ref{fig:uncondgen}d). Although the SA score is only a heuristic measure of synthetic accessibility, this result suggests that the increased valid and novel yield of LlamaUni was not associated with poorer synthetic accessibility scores.

Global and local distributional comparisons further distinguished how the models explore chemical space (Fig.~\ref{fig:uncondgen}e--g). Mean nearest neighbor Tanimoto similarity (SNN) quantifies the similarity of each valid polymer output to its closest PolyInfo polymer. It showed that CharRNN generated structures concentrated near individual PolyInfo polymers, whereas the PolyLatentFlow configurations produced outputs with lower similarity to their nearest PolyInfo neighbors than the other competitive baselines. Fréchet ChemNet Distance (FCD) provided a complementary measure of global distributional agreement, with CharRNN showing the closest agreement with the PolyInfo distribution and the PolyLatentFlow configurations exhibiting greater distributional deviation(Fig.~\ref{fig:uncondgen}e). The joint FCD--SNN analysis therefore illustrates that global agreement with the PolyInfo distribution and proximity to individual PolyInfo polymers capture distinct aspects of chemical space similarity (Fig.~\ref{fig:uncondgen}f). PolyLatentFlow configurations retain moderate agreement while generating structures that are, on average, less similar to their closest PolyInfo neighbors. Furthermore, the UMAP based on Morgan fingerprints provides a qualitative view consistent with this interpretation, with LlamaUni outputs overlapping much of the PolyInfo manifold while also extending into peripheral regions (Fig.~\ref{fig:uncondgen}g). Together with its high validity, novelty relative to PolyInfo, and internal diversity, these results support broader exploration beyond the PolyInfo reference set rather than generation concentrated around a narrow set of reference structures. Full metrics for all models are reported in Supplementary Table S1.

\subsection{Conditional generation for glass transition temperature}

\begin{figure}[t]
\centering
\includegraphics[width=\textwidth]{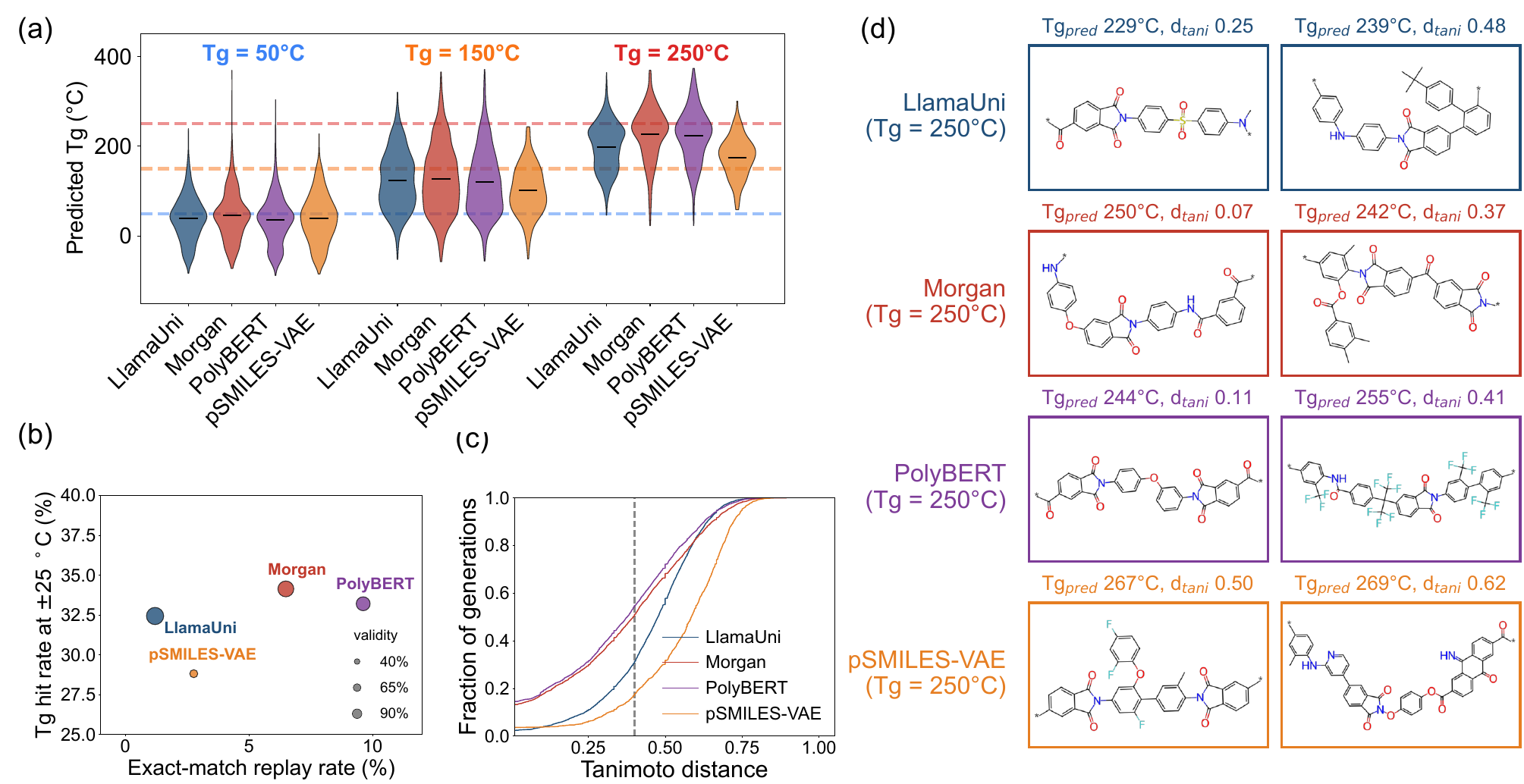}
\caption{\textbf{Similar $T_g$ target responsiveness conceals representation dependent differences in exact-match replay and analogue proximity.} (a) Predicted $T_g$ distributions of valid polymer outputs from 2,000 generation attempts per representation and requested target. Black horizontal marks indicate the means, and colored dashed lines indicate the requested targets. (b) Pooled $T_g$ hit rate within \(\pm25~^\circ\mathrm{C}\) versus exact-match replay rate among valid polymer outputs. Bubble area represents polymer validity across all generation attempts. (c) Cumulative distribution of Tanimoto distance to the nearest polymer in the labeled $T_g$ set. The vertical dashed line marks the analogue novelty threshold of 0.4. (d) Two representative valid polymer structures per representation at a requested \(T_g\) of 250~\(^\circ\mathrm{C}\), selected from within the target window and with an SA score below 5, and annotated with predicted $T_g$ and Tanimoto distance to the nearest labeled polymer. Per target hit and replay rates are reported in Table S2 and Fig.S1.}

\label{fig:condgenTg}
\end{figure}

We next evaluated conditional generation using glass transition temperature ($T_g$) as a single property design task. We generated 2,000 polymers for each representation at target $T_g\in\{50,150,250\}$ °C and scored valid polymer outputs with PolyLLMem predictive model (Fig.~\ref{fig:condgenTg}). The predicted-$T_g$ distributions shifted progressively upward with increasing target for all four representations, demonstrating that each latent space responded to the conditioning signal (Fig.~\ref{fig:condgenTg}a). Although the generated distributions remained below the two higher targets, the consistent monotonic shifts demonstrate that the flow models learned meaningful target-dependent control across a broad temperature range. A complementary UMAP showed corresponding shifts in the LlamaUni output distribution across the labeled polymer space as the requested $T_g$ increased, suggesting that property conditioning was accompanied by systematic changes in generated chemistry (Fig. S2). Predictions from an independently trained Morgan fingerprint random forest were strongly correlated with those from the PolyLLMem ($r=0.897$; Fig. S3), supporting that these conditioning trends were not specific to a single property predictor. Across the three target conditions, pooled hit rates within \(\pm25\,^\circ\mathrm{C}\) were similar across the four representations, remaining near one-third of valid polymer outputs (Table S2). Property fidelity alone therefore provided little basis for distinguishing the representations. Moreover, the 250 °C pSMILES-VAE distribution should be interpreted cautiously because only 113 of 2,000 attempts were polymer valid.

A clearer distinction among representations emerged when target hit rate was considered together with exact reuse of the training set (Fig.~\ref{fig:condgenTg}b). We defined the exact-match replay rate $M$ as the fraction of valid polymer outputs whose canonical pSMILES was identical to a polymer in the corresponding conditional training set. PolyLatentFlow with LlamaUni combined 32.4\% target hit rate with only 1.2\% replay and 93.0\% validity. Morgan and PolyBERT achieved comparable hit rates but replayed 6.5\% and 9.6\% of their valid outputs, respectively, whereas pSMILES-VAE maintained relatively low replay but substantially lower validity. Furthermore, the per-target analysis showed that LlamaUni's low pooled replay was consistent across all three conditioning values, remaining between 0.6\% and 2.1\%, below Morgan and PolyBERT at each target (Table S2 and Fig. S1). LlamaUni's advantage in this task is therefore not higher raw $T_g$ fidelity, but comparable conditional control with consistently less exact reuse of the training set and higher polymer validity.

Distances to the nearest training polymer further quantify how far nonreplayed outputs extend beyond the labeled training set (Fig.~\ref{fig:condgenTg}c). For each valid polymer output, we calculated the minimum Tanimoto distance to the $T_g$ training set using Morgan fingerprints. LlamaUni generated structures farther from the labeled set than Morgan and PolyBERT, with 69.5\% of its valid outputs exceeding the prespecified distance threshold of 0.4, compared with approximately half for Morgan and PolyBERT. pSMILES-VAE generated still more distant valid structures but at substantially lower polymer validity. Differences among representations therefore extend beyond exact replay to the structural proximity of generated polymers to the labeled set, with LlamaUni combining greater separation from labeled chemistry with high polymer validity.

To illustrate the chemical structures associated with these population trends, we selected representative polymers satisfying the $T_g^*=250$ °C target window and an SA score below 5 (Fig.~\ref{fig:condgenTg}d). The Morgan, PolyBERT, and LlamaUni examples include both close analogues of training polymers and more distant candidates, whereas the displayed pSMILES VAE examples are more consistently distant from the labeled set. Across representations, the high $T_g$ candidates frequently contain rigid aromatic and imide motifs while differing in their linkers and substituents.

\subsection{Conditional generation for joint objectives}
\begin{figure}[!htbp]
\centering
\includegraphics[width=\textwidth]{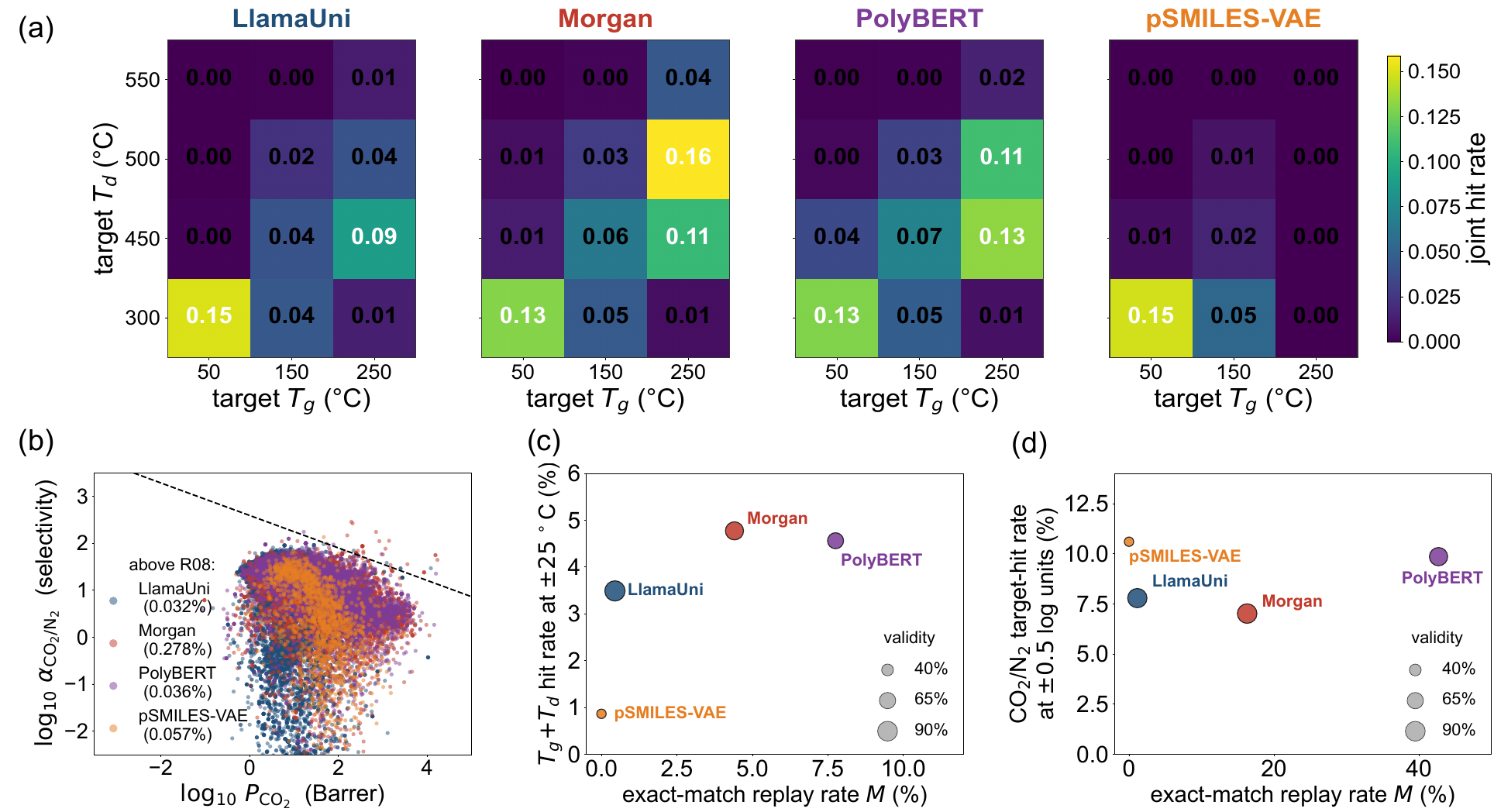}
\caption{\textbf{Coupled property conditioning reveals a representation dependent balance between target fidelity and exact-match replay.}
(a) Joint \(T_g\) and \(T_d\) target hit rate heatmaps using a \(\pm25~^\circ\mathrm{C}\) window across the displayed \(3\times4\) grid of three \(T_g\) targets and four \(T_d\) targets. Cellwise hit rates were calculated among outputs for which both thermal surrogate models returned predictions. (b) Surrogate predicted CO$_2$ permeability and CO$_2$/N$_2$ selectivity for valid polymer outputs with paired predictions. The dashed curve indicates the Robeson 2008 upper bound, \(P_{\mathrm{CO_2}}=30{,}967{,}000\,\alpha_{\mathrm{CO_2/N_2}}^{-2.888}\), with \(P_{\mathrm{CO_2}}\) in Barrer. Legend percentages indicate the fraction of valid polymer outputs predicted to lie above the bound. (c) Joint \(T_g\) and \(T_d\) target hit rate versus exact-match replay rate \(M\), pooled across all 12 target combinations. (d) Joint CO$_2$/N$_2$ target hit rate versus \(M\), pooled across all nine target combinations. A gas hit requires both predicted log permeabilities to fall within \(\pm0.5\) log units of their respective requested values. In (c,d), pooled rates use all valid polymer outputs, and bubble area represents polymer validity across all generation attempts. Per target combination and aggregate conditional metrics are reported in Table S2, and conditional polymer-validity rates for all three regimes are shown in Fig. S4 and Table S4.}

\label{fig:condgenmulti}
\end{figure}

We next extended conditional generation to two-property design tasks with substantially sparser paired supervision. For joint $T_g$ and $T_d$ control, each representation generated 2,000 polymers across a grid of requested property pairs (Fig.~\ref{fig:condgenmulti}a). Joint hit rates varied substantially across the grid, indicating that some combinations of $T_g$ and $T_d$ were more readily achieved than others. All representations showed lower target hit rates at the highest $T_d$ values, consistent with greater difficulty in regions with limited labeled data. These results indicate that PolyLatentFlow learns coupled property dependence, but that achievable control remains strongly influenced by the support of the labeled training data.

We then examined CO$_2$/N$_2$ separation, the most data limited task, with only 623 polymers containing paired permeability measurements. Each representation generated 2,000 samples at each of nine target cells, and valid polymer outputs with finite surrogate predictions were mapped onto the permeability--selectivity plane and compared with the Robeson 2008 upper bound (Fig.~\ref{fig:condgenmulti}b). Predicted exceedances were rare for all four representations, although Morgan produced more such candidates than the other models. Because these exceedances are based on surrogate predictions and do not necessarily satisfy the requested permeability targets, they should be viewed as candidates for higher-fidelity evaluation rather than experimentally validated performance beyond the upper bound.

To evaluate conditional performance more directly, we compared both multi-property tasks in terms of target hit rate and exact-match replay rate $M$. For $T_g$--$T_d$ conditioning, Morgan and PolyBERT achieved only modestly higher joint hit rates than LlamaUni but showed substantially greater exact replay (Fig.~\ref{fig:condgenmulti}c). For CO$_2$/N$_2$, a hit required both predicted log permeabilities to lie within $\pm0.5$ log units of their requested values. The representations showed distinct trade offs between target fidelity, exact replay and polymer validity. pSMILES-VAE achieved the highest hit rate among its polymer valid outputs, but generated valid polymers in only 9.8\% of attempts. PolyBERT also achieved high target fidelity, but with substantially greater exact-match replay.  In contrast, LlamaUni combined low replay with high polymer validity and produced the largest per attempt yield of valid, nonreplayed target hitting candidates. Thus, no representation was uniformly superior, but LlamaUni provided the most robust balance between target attainment and exploration beyond the labeled set.

Because the hit rate and exact-match replay rate were calculated independently, neither directly measures how often a generated polymer both satisfies the requested target and differs from the labeled training set. We therefore defined the nonreplayed target hit rate among valid polymer outputs as

$$J_{\mathrm{valid}}=\frac{N(\mathrm{valid}\cap\mathrm{hit}\cap\mathrm{nonmatch})}{N(\mathrm{valid})}.$$

where a nonmatch is a valid polymer output whose canonical pSMILES is absent from the corresponding corpus with property labels. LlamaUni, Morgan and PolyBERT showed similar nonreplayed target hit rate in the thermal property tasks, but clearer differences emerged for CO$_2$/N$_2$ conditioning (Fig. S5 and table S2). Excluding exact matches to the training set substantially reduced PolyBERT's apparent advantage in raw target fidelity, whereas LlamaUni retained a higher nonreplayed target hit rate. Because \(J_{\mathrm{valid}}\) is calculated only among valid polymer outputs, it does not account for differences in polymer validity across representations. We therefore also calculated the nonreplayed target hit yield per generation attempt,

\[
J_{\mathrm{attempted}}
=
\frac{N(\mathrm{valid}\cap\mathrm{hit}\cap\mathrm{nonmatch})}
{N(\mathrm{attempted})}.
\]
By this measure, LlamaUni produced the largest per-attempt yield of valid polymers that satisfied both property targets and were not exact matches to the training set in the CO$_2$/N$_2$ task (Fig. S5). Overall, no representation was uniformly superior across all conditioning tasks, but LlamaUni consistently combined low replay with high polymer validity and competitive target attainment.

We finally examined how increasing paired supervision affected target hit rate and reuse of the labeled set (Fig. S6). As the CO$_2$/N$_2$ training set increased from $n=100$ to 623, raw target fidelity improved for both LlamaUni and PolyBERT, but the source of this improvement differed substantially. LlamaUni maintained low replay as target hit rate increased, leading to a corresponding increase in nonreplayed target hits. PolyBERT also improved in raw target attainment, but this gain was accompanied by a pronounced increase in exact-match replay, resulting in a much smaller improvement in nonreplayed target attainment (Fig. S6c). Morgan showed intermediate behavior, whereas pSMILES-VAE estimates were less stable because of its low polymer validity. These results indicate that molecular representation shapes how additional supervision translates into chemistry that satisfies the requested targets, either by expanding beyond the labeled set or by increasingly recovering known labeled polymers.

\FloatBarrier

\subsection{Structural exploration and the latent origins of replay}

\begin{figure}[!htbp]
\centering
\includegraphics[width=\textwidth]{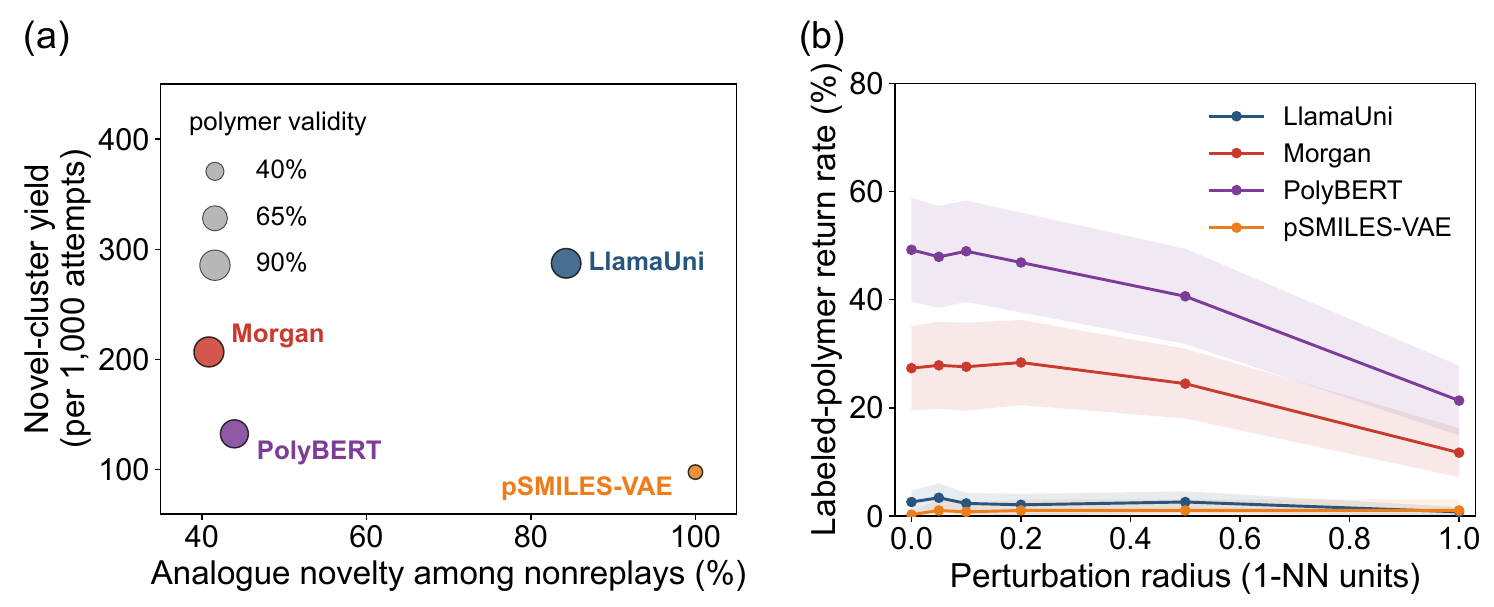}
\caption{\textbf{Structural exploration and decoder capture.}
(a) Analogue novel fraction among nonreplayed valid polymer outputs in the CO$_2$/N$_2$ task versus the yield of distinct analogue novel clusters per 1,000 generation attempts. Bubble area represents polymer validity across all attempts. (b) Probability that stochastic decoding returns a canonical pSMILES matching any polymer in the gas labeled corpus of 623 polymers after perturbing 96 labeled latent representations by the indicated radius. Perturbation radii were normalized by the representation specific median leave one out nearest neighbor spacing among labeled latents. Four stochastic decodes were performed for each source latent and perturbation radius. Shading denotes 95\% normal intervals calculated from standard errors across source latents.}
\label{fig:analysis}
\end{figure}

Exact nonmatch excludes identical training polymers but does not determine whether generated candidates remain close structural analogues of the labeled set. We therefore extended the analysis to scaffold and fingerprint level novelty by comparing Bemis--Murcko scaffolds after removing the two attachment atoms and clustering unique outputs that remained at least 0.4 Tanimoto distant from every polymer in the corresponding labeled corpus (Fig.~\ref{fig:analysis}a and Fig. S7). The CO$_2$/N$_2$ task showed the clearest differences among representations. Among nonreplayed valid polymer outputs, 84.3\% of LlamaUni generations exceeded the analogue distance threshold, compared with approximately 40\% for Morgan and PolyBERT. More importantly, LlamaUni produced the largest yield of distinct analogue novel clusters per generation attempt, whereas pSMILES-VAE showed the highest analogue novel fraction but a much lower cluster yield because it produced fewer valid polymers. LlamaUni also maintained the largest cluster yield in the thermal property tasks (Fig. S7). These results show that its low exact replay is accompanied by broader structural exploration rather than simply avoiding exact matches through small modifications of labeled polymers.

Having established that low exact replay is accompanied by broader structural novelty, we next tested whether decoding itself can contribute to exact replay. We perturbed labeled latent points over increasing standardized radii and repeatedly decoded their neighborhoods with each configuration’s fixed, representation-specific autoregressive decoder (Fig.~\ref{fig:analysis}b). This experiment tests the tendency of a latent--decoder interface to return an exact labeled polymer from nearby latent points. Morgan and especially PolyBERT returned labeled polymers frequently near the training latents, and these returns persisted as the perturbation radius increased. LlamaUni, in contrast, returned labeled polymers only rarely across the same range of perturbations. pSMILES-VAE also showed low return rates, but its much lower polymer validity in this experiment limits direct comparison. These results show that molecular representation influences how broadly latent neighborhoods decode back to known polymers, providing a plausible decoder contribution to the lower replay observed with LlamaUni.

\FloatBarrier

\section{Conclusion}\label{sec:conclusion}

PolyLatentFlow demonstrates that flow matching in latent space provides a flexible and effective framework for unconditional and conditional generation of synthetic polymers. Its multimodal LlamaUni representation produced the largest yield of valid structures novel relative to PolyInfo in unconditional generation and combined responsive conditioning with high validity and consistently low exact replay across three property regimes. In paired CO$_2$/N$_2$ conditioning, pSMILES-VAE achieved the largest target hit rate among its valid polymer outputs but only 9.8\% of attempts were polymer valid, whereas LlamaUni produced the largest per-attempt yield after requiring polymer validity and excluding exact matches to the labeled set. Analogue distance and clustering analyses showed that LlamaUni's low exact replay was accompanied by broader structural exploration rather than only small modifications of labeled polymers. The controlled decoder experiment further showed that molecular representations differ in how broadly local latent neighborhoods decode back to known polymers. Together, these results show that molecular representation shapes whether property conditioning promotes exploration or reuse of labeled chemistry.

These findings also highlight the need to evaluate polymer generators beyond validity and surrogate property fidelity. Exact-match replay, structural distance, and their intersections with target attainment provide complementary measures of whether generated candidates extend beyond the training set. The balance among these criteria should be considered in light of the intended design objective. Concentration around known high performing polymers may be useful for analogue optimization, whereas broader exploration requires low replay, structural novelty, high polymer validity, and target attainment. The data-scaling results further show that additional labeled data do not necessarily increase nonreplayed target hitting yield when improved conditioning is accompanied by greater training set reuse.

Property attainment was evaluated using surrogate models, and prospective simulation, synthesis, and experimental characterization will be required to validate generated candidates as next steps. Exact matching, fingerprint distance, scaffold analysis, chemical clustering and synthetic accessibility scoring offer complementary evidence of how generated structures extend beyond the labeled polymers, while reaction aware assessment will be needed to establish practical synthesis routes. Within this scope, PolyLatentFlow demonstrates the utility of latent-space flow matching for polymer inverse design while revealing molecular representation as a central determinant of the balance between target-property control and exploration beyond labeled chemistry.

\section{Methods}\label{sec11}

\subsection{Data}

PI1M provided the unlabeled pSMILES corpus used for representation compression, unconditional flow training, and autoregressive decoder training \cite{ma2020pi1m}. PolyInfo provided 12,165 canonical pSMILES for the unconditional reference distribution and the thermal datasets with property labels.\cite{otsuka2011polyinfo}. The $T_g$ dataset contained 6,668 polymers, and the paired $T_g$--$T_d$ dataset contained 3,690 polymers. A membrane dataset compiled from literature contained 623 polymers with paired CO$_2$ and N$_2$ permeability measurements.\cite{zhang2025polyllmem}.

\subsection{Molecular representations and compact latents}

LlamaUni combines sequence and three dimensional structural representations of each polymer repeat unit. The pSMILES representation was obtained from a frozen Llama 3 8B model \cite{grattafiori2024llama3} by mean pooling the final hidden states to produce a 4,096 dimensional embedding. The corresponding capped repeat unit dimer was represented using the frozen Uni Mol model \cite{zhou2023unimol}, following the previous work \cite{zhang2025polyllmem}. These two embeddings were concatenated to form the LlamaUni representation. For comparison, Morgan and PolyBERT representations were derived from the same polymer pSMILES without the additional three dimensional structural information. Morgan representations consisted of ECFP fingerprints with 2,048 bits and radius 2 \cite{rogers2010ecfp}. PolyBERT representations were obtained by mean pooling the 600 dimensional final hidden states of the released polymer language model \cite{kuenneth2023polybert}. For LlamaUni, Morgan, and PolyBERT, separate multi-layer perceptron variational autoencoders (VAEs) compressed the corresponding molecular representations into 512-dimensional latent spaces. The posterior mean was used as the compact latent representation for subsequent flow training and decoding. The compact VAEs were trained for 15 epochs on the corresponding PI1M representations using a KL-divergence weight of $10^{-3}$.

The pSMILES-VAE comparator operated directly on tokenized pSMILES. A bidirectional GRU sequence encoder with two layers produced a sequence embedding of fixed length, which a compact VAE compressed into a latent representation with 512 dimensions. Posterior means were used for flow training. All representation models were trained on PI1M. Additional architectural dimensions, optimization settings and procedures for checkpoint selection are provided in Supplementary Methods.

\subsection{Flow matching and decoding}

All flow models operated in the corresponding 512-dimensional compact latent space. Each network contained eight residual blocks with hidden dimension 1,024, layer normalization, SiLU activation, and a 128-dimensional sinusoidal time embedding. Flow matching was performed along straight-line paths between Gaussian noise $z_0$ and data latents $z_1$, $z_t=(1-t)z_0+t z_1$, with velocity target $v(z_t,t)=z_1-z_0$. The flow model was trained to minimize the squared error between the predicted and target velocity fields \cite{lipman2023flow}. Unconditional PI1M flows were trained for 200 epochs, and the checkpoint with the lowest validation flow-matching loss was retained.

For each molecular representation, polymer reconstruction used a separately trained latent-conditioned GRU decoder with the same architecture and training procedure. The decoder contained three GRU layers with hidden dimension 768 and dropout 0.1 and was trained to reconstruct PI1M pSMILES from the corresponding compact latents using teacher forcing and token-level cross-entropy loss. Decoder checkpoint selection considered polymer validity and uniqueness during unconstrained generation. Unless otherwise stated, decoding used a temperature of 0.85 and top-$k=50$. An output was considered polymer valid when it could be parsed by RDKit and contained exactly two attachment atoms defining the repeat unit.

For unconditional generation, the PI1M flow for each representation was fine-tuned on the PolyInfo data set for 100 epochs using a batch size of 256 and AdamW with a learning rate of $5\times10^{-5}$. PI1M latent-normalization statistics and the corresponding PI1M-trained decoder were held fixed. Each model generated 10,000 samples using 100 forward-Euler integration steps and a noise temperature of 0.9. Following inverse latent normalization, each latent coordinate was clipped to $[-5,5]$ before decoding. This clipping represents an absolute bound in latent space rather than clipping at five standard deviations.

\subsection{Conditional training and generation}
Conditional flow models were initialized from pretrained flow checkpoints. The $T_g$ and paired gas models were initialized from their corresponding unconditional checkpoints, whereas the joint $T_g$--$T_d$ models were warm-started from the corresponding $T_g$-conditioned checkpoints. Target properties were standardized using the relevant labeled dataset, and sampling balanced across quantiles was used to reduce imbalance in the property distributions. The checkpoint with the lowest validation flow-matching loss was retained. Classifier-free guidance scales were 1.5 for thermal property generation and 3.0 for gas permeability generation. Unless otherwise stated, conditional sampling used 100 forward-Euler integration steps, a noise temperature of 0.9, coordinatewise clipping to $[-5,5]$ after inverse latent normalization, a decoder temperature of 0.85, and top $k$ sampling with $k=50$. Additional training hyperparameters and details of the validation split are provided in the Supplementary Methods.

For single property $T_g$ generation, 2,000 samples were generated for each representation at target temperatures of 50, 150, and 250 °C. Joint $T_g$--$T_d$ generation used a $3\times4$ target grid with $T_g\in{50,150,250}$ °C and $T_d\in{300,450,500,550}$ °C, with 2,000 generation attempts for each representation and target pair. Gas conditioning used a $3\times3$ grid with physical $\log_{10}(P/\mathrm{Barrer})$ targets of ${0,1.5,3}$ for CO$_2$ and ${-1.5,0,1.5}$ for N$_2$, again with 2,000 attempts for each representation and target pair.

\subsection{Surrogate property prediction}

Each property was predicted with a PolyLLMem model developed specifically for that property\cite{zhang2025polyllmem}. Because the PolyLLMem property predictors and conditional flow models were developed from overlapping property compilations, target hit rates were treated as internal evaluations based on surrogate predictions rather than independent prospective validation. The $J_{\mathrm{valid}}$ and $J_{\mathrm{attempted}}$ metrics described below additionally exclude exact matches to the corresponding corpus with property labels. 

To assess predictor robustness, we compared PolyLLMem \(T_g\) estimates with those from an independently trained Morgan fingerprint random forest for generated polymers that were polymer valid and scoreable by both predictors. Predictions were strongly correlated (Pearson $r=0.897$), with a mean Morgan-RF minus PolyLLMem bias of $+0.5$ °C and per-representation hit rate differences within $\pm3.8$ percentage points (Fig. S3). Thermal decomposition $T_d$ predictions were similarly cross-checked across jointly scoreable outputs from the displayed $3\times4$ grid. The two predictors showed a Pearson correlation of $r=0.725$ and a mean Morgan-RF minus PolyLLMem bias of $+30.3$ °C. Despite this systematic offset, sample-weighted joint $T_g$--$T_d$ hit rates differed by no more than 0.91 percentage points between the two surrogate models (Fig. S8).

\subsection{Evaluation metrics}

Unconditional metrics were calculated on a common 10,000-sample draw. For each published baseline, 10,000 entries were sampled with seed 42 from the pre-generated files released by Yue et al.\cite{yue2025polybench}; the same evaluation pipeline was then applied to baseline and PolyLatentFlow outputs. Uniqueness was defined as the fraction of valid polymer outputs corresponding to distinct canonical pSMILES. Novelty relative to PolyInfo was defined as the fraction of unique valid polymer outputs absent from the PolyInfo reference set. Internal diversity was calculated as the mean Tanimoto distance across 5,000 randomly sampled, nonidentical pairs of unique valid polymer outputs using 2,048-bit radius-2 Morgan fingerprints. Mean nearest-neighbor similarity (SNN) was calculated as the mean maximum Tanimoto similarity between each unique valid polymer generated output and the PolyInfo reference set. Synthetic accessibility was summarized as the fraction of valid polymer outputs with an SA score below 5; this metric was treated as a heuristic and not as evidence of experimental synthesizability. Fréchet ChemNet Distance (FCD) was calculated using \texttt{fcd\_torch} from canonical RDKit-valid outputs relative to the PolyInfo reference distribution \cite{preuer2018fcd,polykovskiy2020moses}.

For conditional generation, exact-match replay was defined as a canonicalized valid polymer output whose pSMILES was identical to a polymer in the corresponding full corpus with property labels used to construct the conditional training and validation subsets. Structural distance from the labeled corpus was quantified as one minus the maximum Morgan fingerprint Tanimoto similarity, using 2,048-bit radius-2 fingerprints. Outputs with a nearest training Tanimoto distance of at least 0.4 were classified as analogue novel according to the prespecified threshold.

For scaffold analysis, the two dummy attachment atoms were removed before calculating Bemis--Murcko scaffolds. Acyclic repeat units were reported separately from scaffold-assessable structures. Unique outputs classified as analogue novel were clustered using RDKit LeaderPicker with a Tanimoto distance cutoff of 0.4, and cluster yield was normalized by the total number of generation attempts. For thermal property conditioning, a $T_g$ or $T_d$ prediction was considered a hit when it lay within $\pm25,^{\circ}$C of the requested target. 

CO$_2$ and N$2$ permeabilities were evaluated on the $\log{10}(P/\mathrm{Barrer})$ scale. Defining $p_i=\log_{10}(P_i/\mathrm{Barrer})$, a gas target was considered attained when both $|\widehat p_{\mathrm{CO_2}}-p^*_{\mathrm{CO_2}}|\leq0.5$ and $|\widehat p_{\mathrm{N_2}}-p^*_{\mathrm{N_2}}|\leq0.5$, where $\widehat{p}_i$ denotes the surrogate-predicted permeability of a generated polymer and $p_i^*$ denotes the requested conditioning target.

Separately, predicted membrane performance was screened against the Robeson 2008 upper bound using $P_{\mathrm{CO_2}}=k\alpha^n$, with $k=30{,}967{,}000$ Barrer and $n=-2.888$\cite{robeson2008upper}. Across regimes, the target hit rate was $H=N(\mathrm{valid}\cap\mathrm{hit})/N(\mathrm{valid})$ and the exact-match replay rate was $M=N(\mathrm{valid}\cap\mathrm{match})/N(\mathrm{valid})$. The nonreplayed target hit rate among valid polymer outputs was $J_{\mathrm{valid}}=N(\mathrm{valid}\cap\mathrm{hit}\cap\mathrm{nonmatch})/N(\mathrm{valid})$, whereas the complementary per-attempt yields $J_{\mathrm{attempted}}=N(\mathrm{valid}\cap\mathrm{hit}\cap\mathrm{nonmatch})/N(\mathrm{attempted})$.

\subsection{Latent analyses and statistics}

To examine whether decoding contributes to exact replay, we assessed decoder capture by perturbing labeled latent representations. Latent dimensions were standardized using the 623 labeled gas polymers, and perturbation radii were normalized by the median leave one out nearest neighbor distance within the labeled latent set. We perturbed 96 labeled latent representations over radii ranging from zero to one normalized nearest neighbor spacing. Four stochastic decodes were performed for each source latent and perturbation radius. A capture event was defined as decoding any canonical polymer present in the labeled gas corpus; invalid decodes were treated as noncaptures. Full perturbation procedures and uncertainty analyses are described in the Supplementary Methods.

To examine the effect of labeled data availability, gas training subsets of 100, 250, and 500 polymers were sampled independently after stratification over a $6\times6$ grid of CO$_2$ and N$_2$ permeability quantile bins; the 623-polymer dataset represented the full-data condition. The trajectories shown in Fig. S6 used 450 generation attempts for each representation and dataset size. Full-data gas models were additionally trained using two different random seeds, and the mean and sample standard deviation across these runs are reported in Table S3.

\section{Data and code availability}\label{sec12}
Source code and a reduced reproduction notebook are available at \url{https://github.com/zhangtr10/polymer_flw}.

\section*{Supplementary information}
Supplementary Methods, Supplementary Figs. S1 to S8, and Supplementary Tables~S1 to S4 are available with this article.

\section*{Acknowledgements}
This work used Jetstream2 at Indiana University through allocation TG-MAT250013 from the Advanced Cyberinfrastructure Coordination Ecosystem: Services \& Support (ACCESS) program, which is supported by U.S. National Science Foundation grants \#2138259, \#2138286, \#2138307, \#2137603, and \#2138296.

\bibliographystyle{unsrtnat}
\bibliography{introduction_references}

\end{document}